\documentclass[english,superscriptaddress,twocolumn]{revtex4}
\usepackage[T1]{fontenc}
\usepackage[latin1]{inputenc}
\usepackage{graphicx}

\makeatletter

\providecommand{\tabularnewline}{\\}

\usepackage{babel}
\makeatother
\begin{document}

\title{Baryon number conservation and enforced electric charge neutrality
for bulk viscosity in quark matter}

\author{Hui Dong}

\affiliation{Department of Modern Physics, University of Science and Technology
of China, Anhui 230026, People's Republic of China}

\author{Nan Su}

\affiliation{Frankfurt Institute for Advanced Studies (FIAS), Max-von-Laue-Str.
1, D-60438 Frankfurt am Main, Germany}

\author{Qun Wang}

\affiliation{Department of Modern Physics, University of Science and Technology
of China, Anhui 230026, People's Republic of China}

\begin{abstract}
General constraints on fluid velocity divergences for particles in
quark matter are derived from baryon number conservation and enforced
electric charge neutrality. A new oscillation pattern in three-flavor
normal quark matter satisfying these conditions is found and its bulk
viscosity is calculated. The result may have astrophysical implication
for maximum rotation frequencies of compact stars. 
\end{abstract}
\maketitle
In the core of compact stars where the density could reach 5-10 times
the normal nuclear matter density, the constituents of nucleons and
hadrons can be squeezed out to form quark matter \cite{Itoh:1970uw,Collins:1974ky}.
Quark matter in normal state could exist in various forms, such as
strangelets \cite{Witten:1984rs,Farhi:1984qu,Schaffner-Bielich:1997fx},
mixed phases \cite{Glendenning:1992vb,Heiselberg:1992dx,Jaikumar:2005ne,Alford:2006bx}
and etc.. Searching for quark matter in stellar objects is very challenging.
The cooling behavior dominated by neutrino emissions may be able to
distinguish nuclear and quark matter \cite{Iwamoto:1980eb,Schafer:2004jp,Alford:2004zr,Jaikumar:2005hy,Schmitt:2005wg,Wang:2006tg,Anglani:2006br}.
But it is still subjected to many experimental and theoretical uncertainties
\cite{Yakovlev:2004iq,Sedrakian:2006mq,Blaschke:2006gd}. Other transport
properties such as shear and bulk viscosities are also of great interest
to this end. The shear viscosity damps the differential rotaion to
make a uniform rigid body rotation of stars. The bulk viscosity is
crucial for the damping of pulsations in compact stars. Such pulsations
could take place during the formation of stars or by external perturbations.
They could also be driven by instabilities of gravitational wave radiation
such as the r-mode instability which arises in the absence of damping
effects. There has been a lot of literature about calculations of
viscosities in nuclear matter in various situations \cite{Itoh:1976,Sawyer:1989dp,Haensel:2000vz,Jones:2001ya,Lindblom:2001hd,Drago:2003wg,Chatterjee:2006hy}.
The shear viscosity for color-flavor-locking phase has been calculated
\cite{Manuel:2004iv}. The bulk viscosities for normal and color superconducting
quark matter have been studied extensively \cite{Wang:1985tg,Sawyer:1989uy,Madsen:1992sx,Zheng:2002jq,Sa'd:2006qv,Alford:2006gy}.
In this letter we put emphasis on general constraints of baryon number
conservation and charge neutrality on the bulk viscosity. Following
these constraints, we will give new solutions to the bulk viscosity
in normal quark matter different from previous results \cite{Wang:1985tg,Sawyer:1989uy,Madsen:1992sx,Anand:1999bj,Zheng:2002jq,Sa'd:2006qv,Alford:2006gy}. 

The bulk viscosity is associated with the damping of the baryon density
oscillation denoted by $\delta n_{B}=\delta n_{B0}e^{i\omega t}$
with the amplitude $\delta n_{B0}$ and the frequency $\omega$ in
the range $10^{3}-10^{4}\;\mathrm{s}^{-1}$. We assume that the amplitude
of the oscillation is small and can be treated as perturbation to
its equilibrium value. The perturbation in baryon density drives quark
matter out of chemical equilibrium via following processes \begin{eqnarray}
u+d & \leftrightarrow & u+s,\;\;(\mathrm{d\leftrightarrow s\; transition}),\nonumber \\
u+e & \leftrightarrow & d+\nu,\;\;(\mathrm{Urca\; I}),\nonumber \\
u+e & \leftrightarrow & s+\nu,\;\;(\mathrm{Urca\; II}).\label{eq:coupled-pr}\end{eqnarray}
Here we consider the normal quark matter with three flavors $u$,
$d$ and $s$. Light quarks $u$ and $d$ are almost massless, while
$s$ quarks have large mass. The relaxation to chemical equilibrium
is related to deviations of chemical potentials from their equilibrium
values for these processes, \begin{eqnarray}
\delta\mu_{1} & \equiv & \mu_{s}-\mu_{d}=\delta\mu_{s}-\delta\mu_{d},\nonumber \\
\delta\mu_{2} & \equiv & \mu_{d}-\mu_{u}-\mu_{e}=\delta\mu_{d}-\delta\mu_{u}-\delta\mu_{e},\nonumber \\
\delta\mu_{3} & \equiv & \mu_{s}-\mu_{u}-\mu_{e}=\delta\mu_{s}-\delta\mu_{u}-\delta\mu_{e}.\label{eq:coupled-chem}\end{eqnarray}
The reaction rates for processes in (\ref{eq:coupled-pr}) can then
be written as $\lambda_{k}\delta\mu_{k}$ ($k=1,2,3$) for small $\delta\mu_{k}\ll T$,
where $\lambda_{k}$ are coefficients. In our calculations, $\lambda_{1}$
and $\lambda_{3}$ are taken from Ref. \cite{Sa'd:2006qv,Alford:2006gy}
except that we have included the mass effect of $s$ quarks \cite{Madsen:1992sx}
and phase space reduction from Fermi liquid behavior \cite{Iwamoto:1980eb,Schafer:2004jp,Wang:2006tg}.
The value of $\lambda_{2}$ is taken from Ref. \cite{Sa'd:2006qv}. 

With massive $s$ quarks a purely three-flavor system cannot be electrically
neutral, so there must be electrons in the system. In reactions (\ref{eq:coupled-pr}),
the electron number and flavors are not conserved since they can be
created or destructed, however the baryon number and electric charge
are conserved. The continuity equations for quark flavor, baryon and
electron number read \begin{equation}
\frac{dn_{i}}{dt}+n_{i}\nabla\cdot\mathbf{v}_{i}=J_{i},\;\;(i=\mathrm{u,d,s,B,e}),\label{eq:cont1}\end{equation}
where the substantial or material derivative is defined by $\frac{d}{dt}\equiv\frac{\partial}{\partial t}+\mathbf{v}_{i}\cdot\nabla.$
The sources $J_{i}$ are linear combinations of reaction rates $\lambda_{k}\delta\mu_{k}$
when $\delta\mu_{k}$ are small. It is obvious that $J_{B}=0$ required
by baryon number conservation. The following relations for baryon
number conservation and charge neutrality are widely used in literature,
\begin{eqnarray}
 &  & n_{B}=\frac{1}{3}\sum_{i=u,d,s}n_{i},\;\; n_{e}=\sum_{i=u,d,s}Q_{i}n_{i},\nonumber \\
 &  & J_{B}=\frac{1}{3}\sum_{i=u,d,s}J_{i}=0,\;\; J_{e}=\sum_{i=u,d,s}Q_{i}J_{i}.\label{eq:neutrality}\end{eqnarray}
The charge neutrality condition in the above is a natural constraint
since the accumulation of net charges would make the Coulomb energy
density blow up and be less favorable in energy. With Eq. (\ref{eq:neutrality})
and continuity equations (\ref{eq:cont1}), one obtains following
constraints on number densities and velocity divergences, \begin{eqnarray}
n_{B}\nabla\cdot\mathbf{v}_{B} & = & \sum_{i=u,d,s}\frac{1}{3}n_{i}\nabla\cdot\mathbf{v}_{i},\nonumber \\
n_{e}\nabla\cdot\mathbf{v}_{e} & = & \sum_{i=u,d,s}Q_{i}n_{i}\nabla\cdot\mathbf{v}_{i},\label{eq:cont3}\end{eqnarray}
where the first line is the result of baryon number conservation and
the second one that of charge neutrality. Eq. (\ref{eq:cont3}) was
not presented and applied explicitly in previous literature. A simple
solution to the above equations can be found at first glance \cite{Wang:1985tg,Sawyer:1989uy,Madsen:1992sx,Zheng:2002jq,Sa'd:2006qv,Alford:2006gy},
\begin{eqnarray}
\nabla\cdot\mathbf{v}_{i} & = & \nabla\cdot\mathbf{v}_{B},\;(i=\mathrm{u,d,s,e}),\label{eq:vivb}\end{eqnarray}
with number densities satisfying Eq. (\ref{eq:neutrality}). This
solution corresponds to the homogeneous or one-component fluid with
a single fluid velocity field. Inserting Eq. (\ref{eq:vivb}) back
into continuity equations (\ref{eq:cont1}), one finds \begin{eqnarray}
n_{B}\frac{dX_{i}}{dt} & = & \frac{dn_{i}}{dt}-X_{i}\frac{dn_{B}}{dt}=J_{i},\label{eq:rate}\end{eqnarray}
where $X_{i}\equiv n_{i}/n_{B}$ are partial fractions for particles
$i$ in terms of the baryon number density. We have used the continuity
equation for the baryon number in Eq. (\ref{eq:cont1}). Obviously
Eq. (\ref{eq:rate}) respects baryon number conservation and charge
neutrality. A property of the solution (\ref{eq:vivb}) is that it
will give a vanishing bulk viscosity in the case that all quarks are
massless, where the system is always in chemical equilibrium. An example
of this property in two-flavor normal quark matter can be found in
Ref. \cite{Sa'd:2006qv}. 

In this paper we rigorously apply the constraints (\ref{eq:cont3})
and find a new solution to Eq. (\ref{eq:cont3}) different from Eq.
(\ref{eq:vivb}). Our starting point is that the role of strange quarks
is special since they are much heavier and may respond to the density
oscillation more reluctantly than other particles. As an extreme case,
we assume $\nabla\cdot\mathbf{v}_{s}=0$. We also assume $\nabla\cdot\mathbf{v}_{u}=\nabla\cdot\mathbf{v}_{d}$.
Then one obtains from Eq. (\ref{eq:cont3}) the following relations,
\begin{eqnarray}
\nabla\cdot\mathbf{v}_{e} & = & \frac{n_{B}(2n_{u}-n_{d})}{n_{e}(n_{u}+n_{d})}\nabla\cdot\mathbf{v}_{B},\nonumber \\
\nabla\cdot\mathbf{v}_{u,d} & = & \frac{3n_{B}}{n_{u}+n_{d}}\nabla\cdot\mathbf{v}_{B}.\label{eq:div-relation}\end{eqnarray}
Generally the velocity divergence for a particle in a fluid depends
on its mass. Rigorously the fluid with more than one particle species
with different masses should be treated as a multi-component fluid.
The above solution to Eq. (\ref{eq:cont3}) is reasonable in the case
that the masses of strange quarks are of the same order as the chemical
potentials and much larger than light quark masses. If the strange
quark masses are small, one can investigate many other solutions which
are close to the solution (\ref{eq:vivb}), for example, one can assume
velocity divergences of particles deviate from that of baryons by
a small amount, $\nabla\cdot\mathbf{v}_{i}=\nabla\cdot\mathbf{v}_{B}+\varepsilon_{i}$
with $(i=\mathrm{u,d,s,e})$. Following Eq. (\ref{eq:cont3}), these
$\varepsilon_{i}$ satisfy $\sum_{i=u,d,s}\frac{1}{3}n_{i}\varepsilon_{i}=0$
and $\sum_{i=u,d,s}Q_{i}n_{i}\varepsilon_{i}=n_{e}\varepsilon_{e}$.
Any small values of $\varepsilon_{i}$ under these constraints denote
a slighly different solution from Eq. (\ref{eq:vivb}). We will not
consider these solutions in this paper and focus on the solution (\ref{eq:div-relation}).
With these relations in Eq. (\ref{eq:div-relation}) for velocity
divergences the bulk viscosity for a system of quarks and electrons
can be derived. The deviation of the pressure from its thermodynamic
value is related to the bulk viscosity \cite{Landau(1987)}, \begin{eqnarray}
\delta\mathcal{P} & = & -\zeta\nabla\cdot\mathbf{v}_{B}.\label{eq:dp1}\end{eqnarray}
The above also provides a definition for the bulk viscosity. The variation
$\delta\mathcal{P}$ can be expressed in terms of density variations
$\delta n_{i}=\delta n_{i0}e^{i\omega t}$ for some quarks or electrons,
which are linearly independent after using Eq. (\ref{eq:neutrality}).
Here $\delta n_{i0}$ are complex amplitudes and can be solved by
applying continuity equations. The number of continuity equations
applied is equal to that of independent densities in order to close
the system of equations. Normally the baryon density $n_{B}$ is set
to be one independent variable. The r.h.s of Eq. (\ref{eq:dp1}) is
actually $\zeta(n_{B})^{-1}dn_{B}/dt$. So a complex $\zeta$ can
be obtained from Eq. (\ref{eq:dp1}), $\zeta=n_{B}\delta\mathcal{P}/(d\delta n_{B}/dt)$,
whose real part gives the bulk viscosity \cite{Landau(1987)}. 

We will first consider in reactions (\ref{eq:coupled-pr}) two simplest
cases, the $d\leftrightarrow s$ transition (by turning off Urca I
and Urca II) and Urca II (by turning off $d\leftrightarrow s$ transition
and Urca I), separately. Finally we will address a more realistic
case, the three coupled processes together. The Urca processes with
light quarks (Urca I) was already studied in Ref. \cite{Sa'd:2006qv}.
The calculations for the $d\leftrightarrow s$ transition without
charge neutrality can be found in Ref. \cite{Alford:2006gy}. We list
all quantities needed in evaluating bulk viscosities in three cases
in Table \ref{cap:quantity}. The second column is for independent
variables. The third one is for continuity equations used in this
paper to solve the variations of densities. For example, we use continuity
equations for $s$ quarks and electrons for the three coupled processes.
With Eq. (\ref{eq:div-relation}) it makes no difference to use other
independent continuity equations. The fourth column lists other variables
expressed in terms of independent ones. In the $d\leftrightarrow s$
transition and Urca II, there are additional constraints on electrons
and $d$ quarks respectively, since they do not participate in the
reactions. With $J_{e,s}=0$ in Eq. (\ref{eq:cont1}) and (\ref{eq:div-relation}),
one can solve $\delta n_{e}$ and $\delta n_{d}$ in terms of $\delta n_{B}$
in the last column. 

\begin{widetext}

\begin{table}

\caption{\label{cap:quantity}Quantities for bulk viscosities in three cases
with $C\equiv3n_{d}/(n_{u}+n_{d})$. }

\begin{tabular}{|c|c|c|c|c|}
\hline 
&
independent variables&
continuity eqs.&
other variables&
special constraints\tabularnewline
\hline 
$d\leftrightarrow s$&
$\delta n_{B}$, $\delta n_{s}$&
$s$&
$\begin{array}{l}
\delta n_{u}=(3-C)\delta n_{B}\\
\delta n_{d}=C\delta n_{B}-\delta n_{s}\end{array}$&
$\begin{array}{c}
J_{e}=0\\
\delta n_{e}=(2-C)\delta n_{B}\end{array}$\tabularnewline
\hline 
Urca II&
$\delta n_{B}$, $\delta n_{s}$&
$s$&
$\begin{array}{l}
\delta n_{u}=(3-C)\delta n_{B}-\delta n_{s}\\
\delta n_{e}=(2-C)\delta n_{B}-\delta n_{s}\end{array}$&
$\begin{array}{c}
J_{d}=0\\
\delta n_{d}=C\delta n_{B}\end{array}$\tabularnewline
\hline 
coupled&
$\delta n_{B}$, $\delta n_{s}$, $\delta n_{e}$&
$s$, $e$&
$\begin{array}{l}
\delta n_{u}=\delta n_{B}+\delta n_{e}\\
\delta n_{d}=2\delta n_{B}-\delta n_{e}-\delta n_{s}\end{array}$&
\tabularnewline
\hline
\end{tabular}
\end{table}

\end{widetext}

\begin{figure}

\caption{\label{cap:zeta1}The bulk viscosity $\zeta_{1}$ for $d\leftrightarrow s$
transition. The masses and chemical potentials (in MeV) are set to
$m_{u}=m_{d}=0$, $m_{s}=100$, $\mu_{s}=\mu_{d}=500$, $\mu_{u}=495$,
$\mu_{e}=5$. Note that these values satisfy charge neutrality and
chemical equilibrium conditions. }

\includegraphics[%
  scale=0.95]{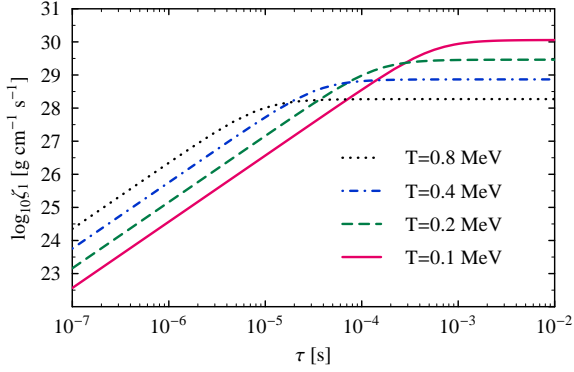}
\end{figure}

For the $d\leftrightarrow s$ transition, following Eq. (\ref{eq:dp1})
and with the second row in Table \ref{cap:quantity}, one obtains
the bulk viscosity, \begin{eqnarray*}
\zeta_{1} & = & \mathrm{Re}\left(\frac{n_{B}\delta\mathcal{P}}{d\delta n_{B}/dt}\right)=\frac{\lambda_{1}n_{B}\frac{\partial\mu_{1}}{\partial n_{B}}\frac{\partial\mathcal{P}}{\partial n_{s}}}{\omega^{2}+\lambda_{1}^{2}\left(\frac{\partial\mu_{1}}{\partial n_{s}}\right)^{2}},\end{eqnarray*}
where \begin{eqnarray*}
\frac{\partial\mu_{1}}{\partial n_{B}} & \equiv & -C\frac{\partial\mu_{d}}{\partial n_{d}},\\
\frac{\partial\mu_{1}}{\partial n_{s}} & \equiv & \frac{\partial\mu_{d}}{\partial n_{d}}+\frac{\partial\mu_{s}}{\partial n_{s}},\\
\frac{\partial\mathcal{P}}{\partial n_{s}} & \equiv & n_{s}\frac{\partial\mu_{s}}{\partial n_{s}}-n_{d}\frac{\partial\mu_{d}}{\partial n_{d}},\end{eqnarray*}
with $C\equiv3n_{d}/(n_{u}+n_{d})$. One can verify that $\zeta_{1}$
is definitely positive since $\frac{\partial\mu_{1}}{\partial n_{B}}<0$
and $\frac{\partial\mathcal{P}_{B}}{\partial n_{s}}<0$ using the
equation of state for degenerate Fermi gas for $d$ and $s$ quarks.
The numerical results for $\zeta_{1}$ are shown in Fig. \ref{cap:zeta1}.
As shown in the figure that the bulk viscosity increases with decreasing
frequency until it satuarates below a critical frequency $\omega_{1c}\sim\lambda_{1}\left|\frac{\partial\mu_{1}}{\partial n_{s}}\right|$.
One also sees that for low frequencies $\omega\ll\omega_{1c}$, $\zeta_{1}$
is inversely proportional to the transport coefficient $\lambda_{1}$.
This means the faster the reaction proceeds the smaller the bulk viscosity
is. 

\begin{figure}

\caption{\label{cap:zeta3}The bulk viscosity $\zeta_{3}$ for the Urca II.
The parameters are the same as in Fig. \ref{cap:zeta1}. }

\includegraphics[%
  scale=0.95]{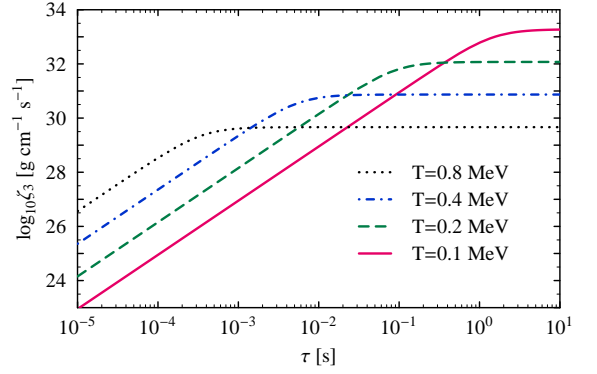}
\end{figure}

Similarly, with the third row in Tab. \ref{cap:quantity}, the bulk
viscosity for Urca II is, \begin{eqnarray}
\zeta_{3} & = & \frac{\lambda_{3}n_{B}\frac{\partial\mu_{3}}{\partial n_{B}}\frac{\partial\mathcal{P}}{\partial n_{s}}}{\omega^{2}+\lambda_{3}^{2}\left(\frac{\partial\mu_{3}}{\partial n_{s}}\right)^{2}},\label{eq:bulk1}\end{eqnarray}
where \begin{eqnarray*}
\frac{\partial\mu_{3}}{\partial n_{B}} & \equiv & -(3-C)\frac{\partial\mu_{u}}{\partial n_{u}}-(2-C)\frac{\partial\mu_{e}}{\partial n_{e}},\\
\frac{\partial\mu_{3}}{\partial n_{s}} & \equiv & \frac{\partial\mu_{s}}{\partial n_{s}}+\frac{\partial\mu_{u}}{\partial n_{u}}+\frac{\partial\mu_{e}}{\partial n_{e}},\\
\frac{\partial\mathcal{P}}{\partial n_{s}} & \equiv & -n_{u}\frac{\partial\mu_{u}}{\partial n_{u}}+n_{s}\frac{\partial\mu_{s}}{\partial n_{s}}-n_{e}\frac{\partial\mu_{e}}{\partial n_{e}}.\end{eqnarray*}
One sees $\frac{\partial\mu_{3}}{\partial n_{B}}<0$ due to $C<2$,
and $\frac{\partial\mathcal{P}}{\partial n_{s}}=-\frac{m_{s}^{2}}{3\mu_{s}}+\frac{m_{u}^{2}}{3\mu_{u}}<0$
with $\mu_{e}\ll\mu_{u}$ and $m_{s}\gg m_{u}$. The numerical results
are shown in Fig. \ref{cap:zeta3}. The behavior of $\zeta_{3}$ is
similar to $\zeta_{1}$. For critical frequencies one sees $\omega_{3c}\sim\lambda_{3}\left|\frac{\partial\mu_{3}}{\partial n_{s}}\right|\ll\omega_{1c}$
because $\lambda_{3}\ll\lambda_{1}$. The saturation value of $\zeta_{3}$
is larger than that of $\zeta_{1}$ for the same reason. At high frequencies
$\omega\gg\omega_{3c}$ one observes $\zeta_{3}\ll\zeta_{1}$. 

\begin{figure}

\caption{\label{cap:zeta}The bulk viscosity $\zeta$ for the coupled processes
at $T=0.1$ MeV. The dashed and dash-dotted curves are the results
from the solution Eq. (\ref{eq:vivb}) with two sets of parameters.
The solid curve is the result from the solution Eq. (\ref{eq:div-relation}).
The parameters of the solid and dashed curves are the same as in Fig.
\ref{cap:zeta1} and \ref{cap:zeta3}, while those of dash-dotted
curve are different and still satisfy neutrality and chemical equilibrium
conditions, $m_{s}=300$, $\mu_{s}=\mu_{d}=500$, $\mu_{u}=456$,
$\mu_{e}=44$, all in unit MeV. }

\includegraphics[%
  scale=0.95]{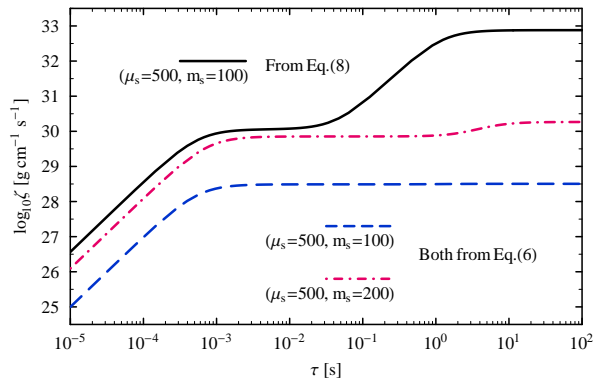}
\end{figure}

With the fourth row in Table \ref{cap:quantity}, the bulk viscosity
for three coupled processes can be expressed as \begin{equation}
\zeta=\frac{n_{B}}{\omega}\left(D_{s}\mathrm{Im}\frac{\delta n_{s}}{\delta n_{B}}+D_{e}\mathrm{Im}\frac{\delta n_{e}}{\delta n_{B}}\right)=\frac{n_{B}}{F}(D_{s}N_{s}+D_{e}N_{e}),\label{eq:zeta}\end{equation}
where $F$ and $N_{e,s}$ are quadratic and linear functions of $\omega^{2}$
respectively. They also depend on coefficients $\lambda_{k}$. The
explicit form of these functions will be presented elsewhere \cite{Dong:2007}.
The coefficients $D_{e,s}$ are given by \begin{eqnarray*}
D_{e} & \equiv & n_{u}\frac{\partial\mu_{u}}{\partial n_{u}}-n_{d}\frac{\partial\mu_{d}}{\partial n_{d}}+n_{e}\frac{\partial\mu_{e}}{\partial n_{e}},\\
D_{s} & \equiv & n_{s}\frac{\partial\mu_{s}}{\partial n_{s}}-n_{d}\frac{\partial\mu_{d}}{\partial n_{d}}.\end{eqnarray*}
 The numerical result for the bulk viscosity $\zeta$ is shown in
Fig. \ref{cap:zeta} by solid curve. Also shown are dashed and dash-dotted
curves from the method with Eq. (\ref{eq:vivb}). With same parameters,
the result from Eq. (\ref{eq:div-relation}) (solid curve) is about
one order of magnitude larger than that from Eq. (\ref{eq:vivb})
(dashed curve). When $m_{s}$ is set to 300 MeV, the result from Eq.
(\ref{eq:vivb}) (dash-dotted curve) is comparable to the solid curve
at high frequencies but still differs from it substantially at low
ones. Generally there can be up to 3 plateaus in the log-log plot
of the bulk viscosity as function of frequency depending on parameters.
In Fig. \ref{cap:zeta}, the solid and dash-dotted curves have two
plateaus, while the dashed curve has only one. It can be verified
that $\zeta$ is dominated by the $s$ quark part in Eq. (\ref{eq:zeta}).
Since $D_{s}$ is negative, the phase of $\delta n_{s}$ should be
delayed relative to $\delta n_{B}$ manifested by the positivity of
the bulk viscosity. The another calculation of the bulk viscosity
for the coupled processes can be found in Ref. \cite{Sa'd:2007ud}. 

In summary, we have derived general constraints on fluid velocity
divergences of quarks and electrons in normal quark matter from baryon
number conservation and enforced electric charge neutrality. Under
these constraints we find a new solution to velocity divergences leading
to different bulk viscosities. As an extreme case, this new solution
could be realistic in some circumstances where strange quarks respond
to the oscillation in a very different way from light quarks. Other
solutions are also allowed by these constraints. The new result for
the coupled processes may have astrophysical implication for larger
maximum rotation frequencies of compact stars. Similar constraints
on fluid velocity divergences can also be obtained for nuclear matter. 

\textbf{Acknowledgement}. The authors thank A.~ Schmitt and I.\ Shovkovy
for valuable discussions, especially A.~ Schmitt for pointing out
our misunderstanding of the method used by some authors. Q.W. acknowledges
support in part by the startup grant from University of Science and
Technology of China (USTC) in association with \emph{100 talents}
project of Chinese Academy of Sciences (CAS) and by National Natural
Science Foundation of China (NSFC) under the grant 10675109.


\begin{thebibliography}{10}
\bibitem{Itoh:1970uw}N.~Itoh, Prog.\ Theor.\ Phys.\ \textbf{44}, 291 (1970). 
\bibitem{Collins:1974ky}J.~C.~Collins and M.~J.~Perry, Phys.\ Rev.\ Lett.\ \textbf{34},
1353 (1975). 


\bibitem{Witten:1984rs}E.~Witten, Phys.\ Rev.\ D \textbf{30}, 272 (1984). 
\bibitem{Farhi:1984qu}E.~Farhi and R.~L.~Jaffe, Phys.\ Rev.\ D \textbf{30}, 2379 (1984).
\bibitem{Schaffner-Bielich:1997fx}J.~Schaffner-Bielich, Nucl.\ Phys.\ A \textbf{639}, 443 (1998)
{[}arXiv:nucl-th/9711044{]}. 
\bibitem{Glendenning:1992vb}N.~K.~Glendenning, Phys.\ Rev.\ D \textbf{46}, 1274 (1992). 
\bibitem{Heiselberg:1992dx}H.~Heiselberg, C.~J.~Pethick and E.~F.~Staubo, Phys.\ Rev.\ Lett.\ \textbf{70},
1355 (1993). 
\bibitem{Jaikumar:2005ne}P.~Jaikumar, S.~Reddy and A.~W.~Steiner, Phys.\ Rev.\ Lett.\ \textbf{96},
041101 (2006) {[}arXiv:nucl-th/0507055{]}. 
\bibitem{Alford:2006bx}M.~G.~Alford, K.~Rajagopal, S.~Reddy and A.~W.~Steiner, Phys.\ Rev.\ D
\textbf{73}, 114016 (2006) {[}arXiv:hep-ph/0604134{]}. 
\bibitem{Iwamoto:1980eb}N.~Iwamoto, Phys.\ Rev.\ Lett.\ \textbf{44}, 1637 (1980). 
\bibitem{Schafer:2004jp}T.~Schafer and K.~Schwenzer, Phys.\ Rev.\ D \textbf{70}, 114037
(2004) {[}arXiv:astro-ph/0410395{]}. 
\bibitem{Alford:2004zr}M.~Alford, P.~Jotwani, C.~Kouvaris, J.~Kundu and K.~Rajagopal,
Phys.\ Rev.\ D \textbf{71}, 114011 (2005) {[}arXiv:astro-ph/0411560{]}.
\bibitem{Jaikumar:2005hy}P.~Jaikumar, C.~D.~Roberts and A.~Sedrakian, Phys.\ Rev.\ C
\textbf{73}, 042801 (2006) {[}arXiv:nucl-th/0509093{]}. 
\bibitem{Schmitt:2005wg}A.~Schmitt, I.~A.~Shovkovy and Q.~Wang, Phys.\ Rev.\ D \textbf{73},
034012 (2006) {[}arXiv:hep-ph/0510347{]}. 
\bibitem{Wang:2006tg}Q.~Wang, Z.~g.~Wang and J.~Wu, Phys.\ Rev.\ D \textbf{74}, 014021
(2006) {[}arXiv:hep-ph/0605092{]}. 
\bibitem{Anglani:2006br}R.~Anglani, G.~Nardulli, M.~Ruggieri and M.~Mannarelli, Phys.\ Rev.\ D
\textbf{74}, 074005 (2006) {[}arXiv:hep-ph/0607341{]}. 
\bibitem{Yakovlev:2004iq}D.~G.~Yakovlev and C.~J.~Pethick, Ann.\ Rev.\ Astron.\ Astrophys.\ \textbf{42},
169 (2004) {[}arXiv:astro-ph/0402143{]}. 
\bibitem{Sedrakian:2006mq}A.~Sedrakian, Prog.\ Part.\ Nucl.\ Phys.\ \textbf{58}, 168 (2007)
{[}arXiv:nucl-th/0601086{]}. 
\bibitem{Blaschke:2006gd}D.~Blaschke and H.~Grigorian, arXiv:astro-ph/0612092. 
\bibitem{Itoh:1976}E.~Flowers and N.~Itoh, Astrophys.\ J.\textbf{\ 206}, 218 (1976). 
\bibitem{Sawyer:1989dp}R.~F.~Sawyer, Phys.\ Rev.\ D \textbf{39}, 3804 (1989). 


\bibitem{Jones:2001ya}P.~B.~Jones, Phys.\ Rev.\ D \textbf{64}, 084003 (2001). 


\bibitem{Lindblom:2001hd}L.~Lindblom and B.~J.~Owen, Phys.\ Rev.\ D \textbf{65}, 063006
(2002) {[}arXiv:astro-ph/0110558{]}. 
\bibitem{Haensel:2000vz}P.~Haensel, K.~P.~Levenfish and D.~G.~Yakovlev, Astron.\ Astrophys.\textbf{\ 357},
1157 (2000) {[}arXiv:astro-ph/0004183{]}. 
\bibitem{Drago:2003wg}A.~Drago, A.~Lavagno and G.~Pagliara, Phys.\ Rev.\ D \textbf{71},
103004 (2005) {[}arXiv:astro-ph/0312009{]}. 
\bibitem{Chatterjee:2006hy}D.~Chatterjee and D.~Bandyopadhyay, Phys.\ Rev.\ D \textbf{74},
023003 (2006) {[}arXiv:astro-ph/0602538{]}. 
\bibitem{Manuel:2004iv}C.~Manuel, A.~Dobado and F.~J.~Llanes-Estrada, JHEP \textbf{0509},
076 (2005) {[}arXiv:hep-ph/0406058{]}. 
\bibitem{Wang:1985tg}Q.~D.~Wang and T.~Lu, Phys.\ Lett.\ B \textbf{148}, 211 (1984).
\bibitem{Sawyer:1989uy}R.~F.~Sawyer, Phys.\ Lett.\ B \textbf{233}, 412 (1989) {[}Erratum-ibid.\ B
\textbf{237}, 605 (1990){]}. 
\bibitem{Madsen:1992sx}J.~Madsen, Phys.\ Rev.\ D \textbf{46} (1992) 3290. 
\bibitem{Anand:1999bj}J.~D.~Anand, N.~Chandrika Devi, V.~K.~Gupta and S.~Singh, Pramana
\textbf{54}, 737 (2000). 
\bibitem{Zheng:2002jq}X.~p.~Zheng, S.~h.~Yang and J.~r.~Li, Phys.\ Lett.\ B \textbf{548},
29 (2002) {[}arXiv:hep-ph/0206187{]}. 
\bibitem{Sa'd:2006qv}B.~A.~Sa'd, I.~A.~Shovkovy and D.~H.~Rischke, arXiv:astro-ph/0607643.
\bibitem{Alford:2006gy}M.~G.~Alford and A.~Schmitt, arXiv:nucl-th/0608019. 
\bibitem{Landau(1987)}L.~D.~Landau and E.~M.~Lifshitz, \emph{Fluid Mechanics, Course
of Theoretical Physics, Volume 6}, Butterworth Heinemann, 1987. 
\bibitem{Dong:2007}H.~Dong, N.~Su and Q.~Wang, work in progress. 
\bibitem{Sa'd:2007ud}B.~A.~Sa'd, I.~A.~Shovkovy and D.~H.~Rischke, arXiv:astro-ph/0703016.
\end{thebibliography}
\end{document}